\documentclass[reprint,
 amsmath,amssymb,aps,
prd,
]{revtex4-1}
\usepackage{graphicx,subfigure,color} 
\usepackage{dcolumn}% Align table columns on decimal point
\usepackage{bm}% bold math
\usepackage{hyperref}% add hypertext capabilities
\usepackage[mathlines]{lineno}% Enable numbering of text and display math

\newcommand{\be}{\begin{equation}}
\newcommand{\ee}{\end{equation}}
\newcommand{\bea}{\begin{eqnarray}}
\newcommand{\eea}{\end{eqnarray}}
\newcommand{\vev}[1]{\langle #1 \rangle}

\begin{document}
%\preprint{APS/123-QED}

\title{\Large Bridging  the Chiral symmetry and Confinement with  Singularity}
\author{Yoon-Seok Choun}
\author{Sang-Jin Sin}
\email{ychoun@gmail.com}
\email{sangjin.sin@gmail.com}
\affiliation{Department of Physics, Hanyang University, 222 Wangshimni-ro, Sungdong-gu, Seoul, 04763, South Korea }
\begin{abstract} 
We consider a holographic quark model where the confinement is a consequence of the quark condensate. Surprisingly, the equation of motion of our holographic model can be mapped to the old spin-less bag model. Both models correctly reproduce the linear Regge trajectory of hadrons for zero quark mass. For the case of non-zero quark mass, the model lead us to Heun's equation. The mass term is precisely the origin of the higher singularity, which changes the system behavior drastically. Our result can shed some light on why the chiral transition is so close to the confinement transition. In the massive case, the Schroedinger equation is exactly solvable, but only if a surprising new quantization condition, additional to the energy quantization, is applied.
\end{abstract}
\date{\today}% 

%\arxivnumber{}

  
\maketitle

%\section
{\bf 1. Introduction: }
A frequent question for the phase diagram of the quantum chromo dynamics (QCD) \cite{Shuryak:2004pry,Yagi:2005yb} is why the chiral  and the de-confinement transitions are so close, while  two are   separate concepts.   The former is  defined when  the current quark mass is zero,  
while the latter is  due to the infrared   dynamics  QCD which is  often summarized by the  QCD string 
\cite{Mandelstam:1974pi,tHooft:1975yol} 
and its spectrum called Regge trajectory, 
 \be \alpha ' m^2=n+\beta. \ee   
The  (approximate)  coincidence of the phase boundaries  will be explained if one can show that  one is a consequence of the other. 

Some time ago,    G\"ursey \cite{Gursey}  showed that  the spectrum 
of  semi-classical bag model  introduced by Lichtenberg et.al\cite{Lich1982} for the meson follows the Regge trajectory if the quark mass vanishes. 
In a recent paper \cite{Bag2019},  the model for non-vanishing quark mass was studied numerically and the spectrum turns out to be highly non-linear.  Since the linear confinement   appears  only for the vanishing  quark mass,  which allows to define the chiral symmetry, 
 one may wonder if   the chiral symmetry   is a consequence of the  the confinement dynamics.   
   
In this paper, we consider  a holographic  model which  gives the linear confinement 
in the presence of the quark condensate, so that the confinement is consequence of the quark condensate in this model.  This is a holographic fermion model coupled with a neutral scalar.   We show   that the equation of motion of this  model    can be mapped to the  Lichtenberg model. 
Considering that two models are based on completely different idea, presence of such mapping is quite remarkable.  In both models the presence of current quark mass is inconsistent with Regge trajectory, and  the quark mass triggers the change of the singularity type from the hypergeometric  type to Heun's type.  The linear trajectory  appears at a limit   where higher singularity disappears.

  We will develop  polynomials whose roots gives quantized value of hadron spectrum and we explicitly calculated the  hadron spectrum  in the presence of the quark mass to understand the spectrum analytically. 
It turns out that  the drastic change of  spectrum in the presence of quark mass is a consequence of change of the singularity type, which requests  an extra quantization: apart from the energy, one more parameter in the potential should be quantized, which is rather a surprising phenomena.   
One should notice that this  is  relevant to   general situations:  whenever  Schr\"odinger equation has the potential with both even and odd powers  of radial coordinate, there  are extra quantizations apart from that of energy.   

Finally, we also emphasize that  the massless limit of the spectrum is singular.  Such  inconsistencies of the spectrum of hadron in the presence of the quark mass suggests that the chiral symmetry should be tied with the color confinement, although the dynamical mechanism of  suppressing the quark mass is still an open question.

 \vskip 0.3cm
 %\section
{ \bf 2. Holographic fermion as a constituent quark:}
To consider the hadron mass problem in terms of effective theory, it is   convenient to consider a model of constituent quark, where 
all the correlation by the gluons are encoded into the constituent  quark mass. Namely, we consider a fermion $\chi$  in a bag which is dual to the fermion $\psi$ living outside the central region of the AdS. The   dynamics of   $\psi$ in the warped space  determines the mass of the excitation, which  we interpret as  the  constituent quark. Such mass of the constituent quark   can be used to describe  meson mass as well as   baryon mass, by assuming that there is no interaction between constituent quarks.  
  
For this, we consider following  fermion action in AdS space with coupling to the scalar describing the bare quark mass and chiral condensation $M$.  
 \bea
 S_{\psi}&=& \int d^{d+1}x\sqrt{-g}
 i\bar{\psi} \big(\Gamma^\mu\mathcal{D}_\mu-m-\Phi  \big)\psi +S_{\Phi},  \label{DiracS}
\eea
 where  ${\cal D}_\mu = \partial_{\mu} +\frac{1}{4} \omega_{ab\mu}\Gamma^{ab}$. 
 One may simply consider this as a model for a Baryon instead of a constituent quark. 
We consider only $d=3$ for the analytical simplicity. 
The dynamics of the boson $\Phi$ is given by  \begin{eqnarray}
S_{\Phi}= \int d^{d+1}x\sqrt{-g}\Big( 
%-\frac14 F_{\mu\nu}^2 
 -|\partial_\mu\Phi |^2 -m_\Phi^2 |\Phi|^2 \Big), \label{action}
\end{eqnarray}
% where  ${D}_\mu= \nabla_{\mu}  -i qA_\mu $ is the covariant derivative.  
We  treat all the fields in the probe limit where 
 the metric is fixed  as that of AdS$_{4}$: 
\be
ds^{2}=(dz^{2}+\eta^{\mu\nu}dx^{u}dx^{\nu} )/z^{2},  \hbox{ with } \eta^{00}=-1.
\ee
Bulk mass of the boson, $m_{\Phi}^2$, is given in terms of the conformal dimension of the dual operator: $m_{\Phi}^2 =\Delta(\Delta-d)$. We will fix it such that $\Delta=2$, so that
 $m_{\Phi}^2 =-2$  in  $d=2+1$ and $m_\Phi^2=-4$ for $d=3+1$.   Although $\Delta=2$ for the operator ${\bar q q}$   is  realized  in 4 dimension at the lower boundary of conformal window of  $N_{f}/N_{c}$ \cite{Kaplan:2009kr}, 
here we consider 2+1   case only. 
The    field equation then  gives 
\bea
\Phi =M_0 z + {M}z^2,  \hbox{ in AdS$_{4}$, } \label{Phi}
%\Phi &=& M_0 z^{2}\ln z^{-1}+ {M}z^2,  \hbox{ in AdS$_{5}$. } \label{Phi}
\eea
which is an   exact solution of the scalar field equation {\it in the probe limit}.
%In  the case  $M_{0}=0$ case was considered.
%In this paper, we consider   general  $M_{0}\neq 0$ case where source is also included. 
%%We turn off the Maxwell field in our analysis. 
%The equation then is given by
%$\nabla^{\mu}F_{\mu\nu}=J_{\nu}$ 
%and for the real solution of $\Phi$, the current is simplified to
%the London equation similarly to the superconductivity,
%$
%J_{\mu}=  \Phi ^{2}A_{\mu}.
%$
%For the transverse components with $\vec{k}\cdot \vec{A=0$, 

  The equation  of motion of
 (\ref{DiracS}) is given by 
 \be
 \big(\Gamma^\mu\mathcal{D}_\mu-m-q\Phi  \big)\psi =0,
 \ee
which    can be written as a Schr\"odinger equation 
%\cite{Karch:2006pv} 
\bea
-\Psi''_{n}(z) &+&V(z)\Psi_{n}(z)=E_{n}\Psi_{n}(z), \qquad \label{vectorE} \\
\hbox{with \qquad } 
	V(z)&=&\frac{m(m-1) +\Phi^{2}}{z^2},  \\
	&=&\frac{m(m-1)}{z^2} +   q^{2}(M z +M_{0})^{2}.\\
	E_n&:=& m_{n}^{2}- 2qM(m+\frac{1}{2})
	\eea
We interpret $m_{n}^{2}$ as the constituent quark mass inside a Hadron   and it was shown that for  $M_{0}=0$,   spectrum is linear \cite{Oh:2019zbr}  
	\be
	m_{n}^{2}=4 qM(n+m+1/2).
	\ee
Notice that when $M=0$ we have 
$m_{n}^{2}=\omega^2-k^2=0$ so that  the only spectrum is massless one with   $\omega=k$, which can not be the spectrum of a confined object:   the energy of the   confined massless quark contribute to the mass of the hadron containing it.  That is what we mean by constituent quark mass. Therefore we can say that $M$ is   the order parameter of the  confinement transition  as well as  that of the chiral transition. Notice that in 2+1 dimension there is no chiral symmetry. 
By chiral symmetry breaking(CSB)  in this paper, 
what we meant is  'non-zero quark condensate' $M$, which is equivalent to CSB in 3+1 dimensional case. The chiral symmetry itself is not relevant to our discussion.

We also   emphasize  that since  $M\sim \vev{\bar q q}$ is the slope of the Regge trajectory,  
the linear confinement is consequence of the non-zero quark condensate.
Therefore  two  transitions   must be  identical  in this model.  Although it is not clear whether  this is a model specific property  or  a generic property of QCD, above argument  explain the coincidence of two transition at least partially in the context of  this specific model.

When the quark mass  $M_{0}\neq 0$,   we will show shortly that  it will lead to a type of Heun's equation.

\vskip.3cm
%\section
{\bf 3. Heun's equation}: 
If we formally replace $ z\to r$,   
\be  \;  m\to   -L,   \; 
qM_{0}\to m_{q},   \; qM\to \frac{b}2,  \hbox { and } \Psi_{n}\to u , \label{corresp}
\ee 
%$E_{n}\to E^{2}/4 $ 
 then eq. (\ref{vectorE})  defined in AdS4 space becomes 
\bea
\left[ -\frac{d^{2}}{dr^{2}} +V(r)  \right]u(r) &=& E u(r), \\
%\hbox{ with }
 V(r)=\Big(m_{q}+ \frac{1}{2}b_{\tau} r\Big)^2 &+&  \frac{L(L+1)}{r^2} ,\label{heun} 
\eea
which is a   Heun's equation\cite{NIST,Ronv1995,Slavy2000} with 4-singularies.
%$\vec{\mathbf{p}}^2=  $ with $P_r = -i \frac{ 1}{r}\frac{\partial}{\partial r}r $. 
One interesting observation is that   above equation    is  precisely the radial equation  coming from the bag model   \cite{Lich1982,Gursey,Bag2019} for a meson, whose mass squared is given by   $4E$. 
We emphasize that the physical ideas and the spaces in which they are defined are completely different: one in AdS$_4$ and the other in  a flat space ${\bf R}^{3}$.

  To reveal  the mathematical structure more clearly,  we consider slightly   generalized  one defined by the potential 
 \be 
 V(r) = c^2 r^2 + b r -\frac{a}{r}+ \frac{L(L+1)}{r^2}, \label{Pabc}
 \ee
 which is obtained from  the potential eq.(\ref{heun}) by shifting 
 $V\to V-a/r -m_{q}^{2}$ and redefining $c =b_{\tau}/2$ and $ b=m_{q}b_{\tau}$. 
 Factoring out the behavior near $r=0$ by  $u(r)= r^{L+1} f(r)$,   above equation becomes
 \begin{equation}
 \frac{d^2 f( {r})}{d{ {r}}^2}  +  \frac{2(L+1)}{ {r}} \frac{d f( {r})}{d {r}} + \left(  {E} - {c}^2  {r}^2-  {b} {  r}
 +  \frac{ {a}}{ {r}} \right) f( {r}) = 0.
\label{qq:4}
\end{equation} 
Factoring out    near   $ \infty$ behavior by  
$ f( {r}) = \exp\left( -\frac{ {c}}{2} {r}^2-\frac{ {b}}{2 {c}} {r}\right)y( {r})$  and 
introducing   $\rho = \sqrt{ {c}} {r}$, $ a_0 = { {a}}/{   {c}^{1/2}}$, 
$b_0 = { {b}}/{ {c}^{3/2}}$,  $\mathcal{E}=  { {E}}/{ {c}}$,
we get   the  bi- confluent Heun's   equation: 
\begin{equation}
\rho  \frac{d^2{y}}{d{\rho}^2} + \left( \mu \rho^2 + \varepsilon \rho + \nu  \right) \frac{d{y}}{d{\rho}} + \left( \Omega \rho + \varepsilon \omega \right) y = 0.
\label{eq:1}
\end{equation}
with 
%\bea \label{Hol}
%& \rho \frac{d^2 y}{d\rho^2}  + \left(-2\rho^2 - b_0 \rho + 2(L+1) \right) \frac{d y}{d\rho} + \\
% & \left[
%  \left( \mathcal{E}  + b_0 ^2/4 -(2L+3) \right) \rho +  a_0  -b_0 (L+1)  \right] y(\rho) = 0 , \nn
%\eea
% whose canonical form is
 $\mu=-2$, $\varepsilon=-b_{0}$, $\nu=2L+2 $
 and 
 \be \Omega= \mathcal{E}  + b_0 ^2/4 -(2L+3), \hbox{ and }
 \omega=L+1-a_{0}/b_{0}.
 \ee
  It has a regular singularity at the origin and an irregular singularity  of rank  two \cite{NIST,Ronv1995,Slavy2000} at the infinity.
 
Substituting $y(\rho)= \sum_{n=0}^{\infty } d_n \rho^{n}$ into (\ref{eq:1}), we obtain the   recurrence relation:
\bea
d_{n+1}&=&A_n \;d_n +B_n \;d_{n-1}  \quad  
\hbox{  for  } n \geq 1, \label{eq:3} \hbox{ with } \\
 A_n&=&-\frac{\varepsilon (n+\omega  )}{(n+1 )(n+\nu )},  \quad   B_n=-\frac{\Omega +\mu (n-1 )}{(n+1 )(n+\nu )}.\eea
  The first two $d_{n}$'s are given by   $d_1= A_0 d_0$ and $d_{-1}=0$.
%Comparing (\ref{Hol}), (\ref{eq:1}),  the former is a special case of the latter with  $\mu = -2$, $\varepsilon = - b_0,  \nu  =  2L+2$ and
%  \bea
%\omega = L+1 -\frac{ a_0 }{ b_0 }, \quad  \Omega  = \mathcal{E}+\frac{b_0 ^2}{4}   -(2L+3). \label{Omega}
%\eea

It is essential to notice that 
when $m_{q}=0$, we have 
\be A_{n}=\varepsilon=b_{0}=0,\ee
 so that
the three term recurrence relation given in eq. (\ref{eq:3})
is reduced to two term recurrence relation between $d_{n+1}$ and $d_{n-1}$ and the Heun's equation is reduced to hypergeometric one.  That is, the quark mass   is precisely the term increasing  the singularity order. 

\begin{figure}[!htb]
 \centering
  \subfigure[${\cal P}_{4}(a_{0},b_{0})=0$]
  { \includegraphics[width=0.47\linewidth]{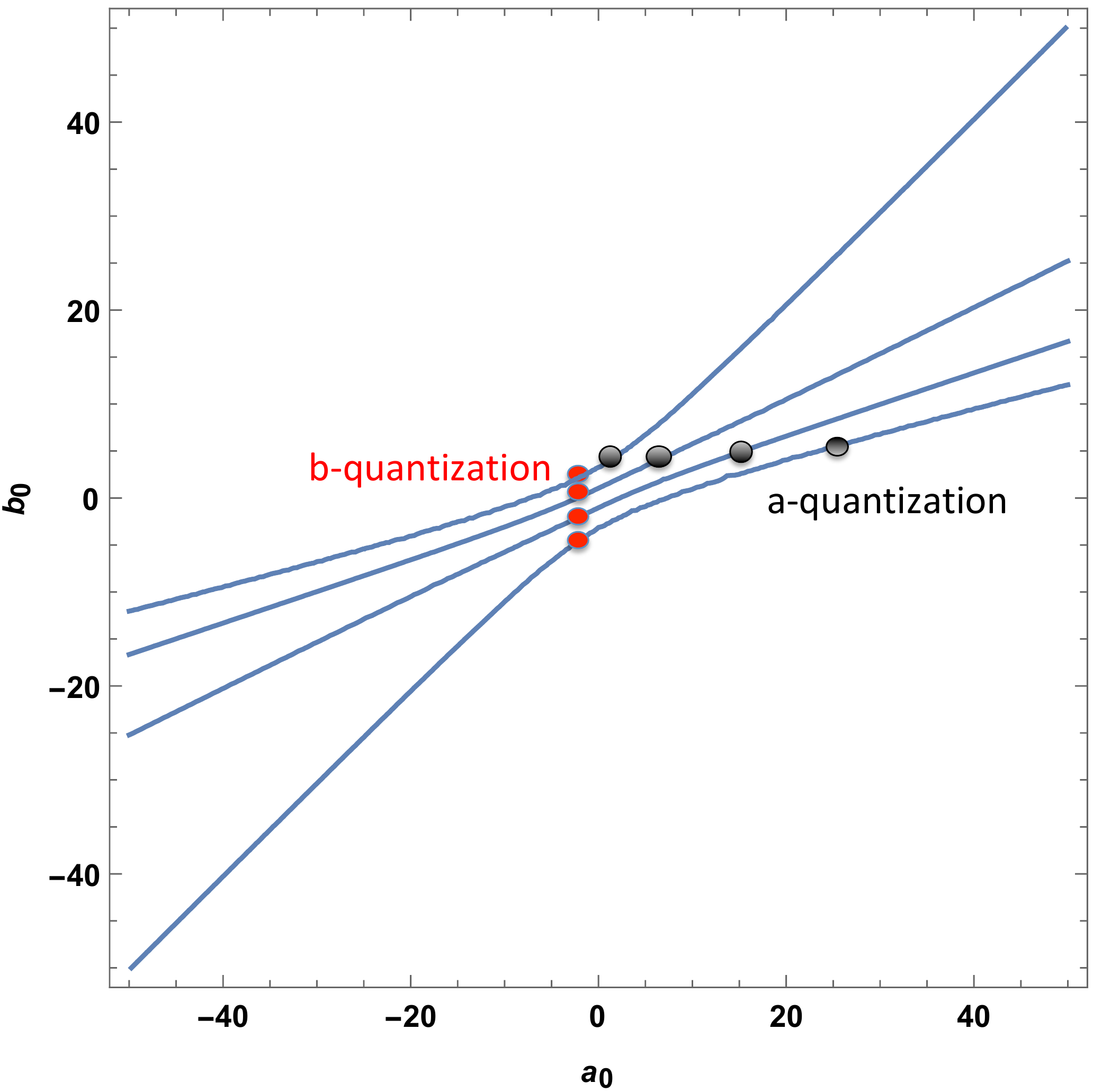}}
 \subfigure[asymptotic view of (a)]
 { \includegraphics[width=0.47\linewidth]{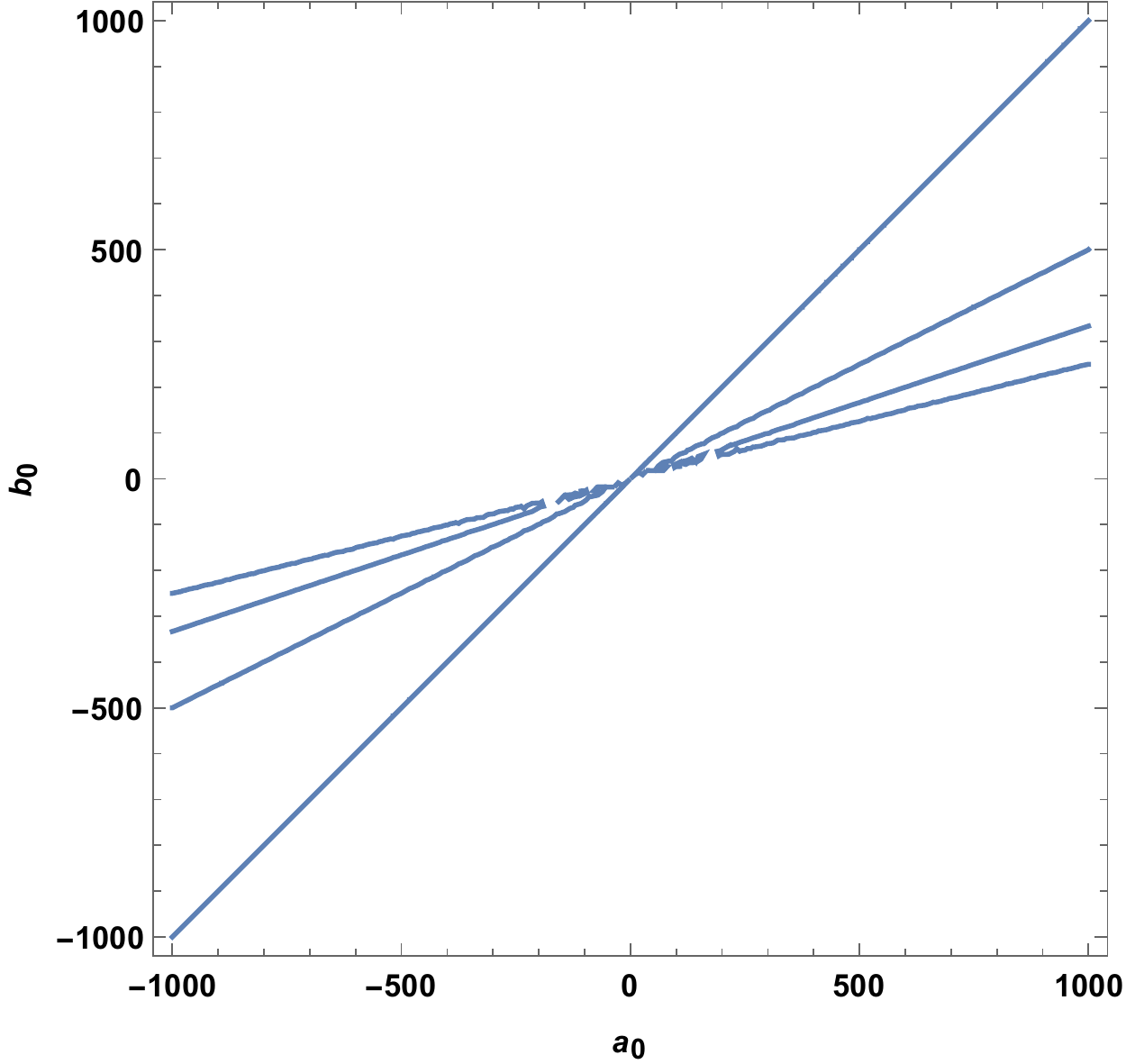}
}   \caption{Contour of ${\cal P}_{4}(a_{0},b_{0})=0$ and its asymptotic view.  Fig. (a) also shows definition of $a_0$(or $b_0$)-quantization depending on we fix $b_0$ or $a_0$. }  \label{Root1}
\end{figure}

Now, unless  $y(\rho)$ is a polynomial, $u(r)$ is divergent as $ \rho\rightarrow \infty$. Therefore we need to impose regularity conditions by which the solution is normalizable.
If we  impose two conditions \cite{NIST,Ronv1995,Slavy2000},
\begin{equation}
B_{N+1}= d_{N+1}=0\hspace{1cm}\mathrm{where}\;N\in \mathbb{N}_{0}, 
 \label{bb:2}
\end{equation}
the series expansion  becomes a polynomial of degree $N$:  
as one can see from eq. (\ref{eq:3}), 
eq. (\ref{bb:2}) is sufficient to give  $d_{N+2}=d_{N+3}=\cdots=0$ recursively. Then   the solution is a polynomial of order $N$, 
$
  y_{N}(\rho)=  \sum_{i=0}^{N}d_{i}\rho^{i}.
 $
The question whether imposing both  equations in eq(\ref{bb:2}) is really necessary     was studied numerically in our earlier work \cite{Bag2019}.
In general, $d_{N+1}=0$ will define a $N+1$-th order polynomial  $ {\cal P}_{N+1} $  in $a_{0},b_{0}$, so that Eq. (\ref{bb:2}) gives 
 \be
\mathcal{E}_{N,L}=2N+2L+3 -  {b_0 ^2}/{4} , \quad 
{\cal P}_{N+1}(a_{0},b_{0})=0.  \label{Omega}
\ee
where the first   comes from $B_{N+1}=0$,  and it is nothing but the usual energy quantization condition. 
Below we will examine the meaning of  the second equation. To do that we need explicit expressions of  a few lower     order  polynomial $ {\cal P}_{N+1}$: 
{\scriptsize
%\footnotesize 
\bea 
\label{app:1}
\begin{split} {\cal P}_{1}(a_{0},b_{0})&=  {b_0 (L+1)-a_0 },\\
 {\cal P}_{2}(a_{0},b_{0})&=   {\left(b_0 (L+1)-a_0 )(b_0 (L+2)-a_0 \right)}- 4(L+1)   ,\\
 {\cal P}_{3}(a_{0},b_{0})&=(L+1)(L+2)(L+3) b_0^3 -(3L(L+4)+11)a_0 b_0^2 \\
&+ \Big(3(L+2)a_0^2 -4(L+1)(4L+9)\Big)b_0  -a_0^3 + 4(4L+5) a_0  , \\
{\cal P}_{4}(a_{0},b_{0}) & = (L+1)(L+2)(L+3)(L+4) b_0^4 -2(2L+5)(L(L+5)+5)a_0 b_0^3 \\
&+ \Big( (6L(L+5)+35)a_0^2 -4(L+1)(5L(2L+11))+72 \Big)b_0^2 \\
 & - \Big(2(2L+5)a_0^3 +4(20L(L+4)+69)a_0  \Big) b_0 -20(2L+3) a_0^2 \\
 &+144(L+1)(L+2)+a_0^4 ,
\end{split}
\eea
}
%\section
{\bf 4. Extra Quantization} : 
We  have seen  that  $a_{0},b_{0}$  should be related by ${\cal P}_{N+1}(a_{0},b_{0})=0$. 
   This means that if we fix one of them, 
   the other should be a solution of a polynomial equation. 
Let's examine a few low orders in N.
 We normalize the solution using $d_{0}=1$ for simplicity.
\begin{enumerate}
\item   For $N=0$: 
${\cal P}_{1}(a_{0},b_{0})=  {b_0 (L+1)-a_0 }  =0$. 
The  eigenfunction is $y_{0}(\rho) =1$.
  \item For $N=1$,   $  {\cal P}_{2}(a_{0},b_{0})=0 $  defines a hyperbola in $a_{0}, b_{0}$ such  that there are always two branches because the discriminant   $D=b_0 ^2+16(L+1) >0$.  That is,  for a  given $b_{0}$,   $ a_0$ always has real solutions. 
  $ 2a_0 = {b_0 (2L+3)\pm \sqrt{b_0 ^2+16(L+1)}} $.
In this case, $y_1(\rho)= 1+d_{1}\rho$  with   $d_{1}= \Big({-b_0  \pm \sqrt{b_0 ^2+16(L+1)}} \Big)/{(4L+4)}  $.

\item For $2\leq N$,    
${\cal P}_{N}=0$ has $N$ branches.  ${\cal P}_{4}=0$
 is plotted in figure~\ref{Root1}.   Apart from the central region where $a_{0},b_{0}\sim {\cal O}(1)$ which is shown in figure \ref{Root1}(a) the curves are  approximately linear.
Such linearity can be confirmed by drawing the same figure  in large scale  as  in figure~\ref{Root1}(b), where we used $L=0$.
Note that the slopes of the lines: $a_{0}/b_{0}=1,2,3,4$.
%It is also worthwhile to note that the medium curve with asymptotic slope 1/2 pass through $a_{0}=0, b_{0}=0$. This happens for all even integer $N$.
%See figure~\ref{Root3} and figure~\ref{Root4}, where   we also used $L=0$ and the slope of the lines can be read off to give $b_{0}/a_{0}=1,1/2, 1/3,1/4$.
\end{enumerate}
 For general $N$,  we can show that for large enough $a_{0},b_{0}$,  ${\cal P}_{ㅜ+1}(a_{0},b_{0})=0$ gives following relation.
\begin{equation}
 \frac{a_0}{b_0} = \frac{ {a} {c}}{ {b}} \simeq L+1+K, \hbox{ for }  K=0,1,\cdots,N.
\label{quanti}
\end{equation}
%where  $0\leq L \leq N$.
 This means that   for a given $b_{0}$, there are $N+1$ $a_{0}$'s  for any $L$.  This is also true for   $|a_{0}|,|b_{0}|\leq {\cal O}(1)$ although we can not write down the explicitly. Similarly
 if we set $ a_0 =0$,    the allowed values of $ b_0 $ are given by the crossing points of $N+1$ branches  of the ${\cal P}_{N+1}=0$ with the vertical line $a_{0}=0$. We call such fixing $b$-quantization. See figure \ref{qz}.

Such extra quantization is a consequence of the Heun's   equation. 
As we have seen before explicitly,    for the hypergeometric equations,  the the three term 
   recurrence relation is reduced to two term  one after factoring out the asymptotic form so that   we need to fine tune only one parameter, the energy,  to have a polynomial  solution.   For the Heun's equation,  its higher
   singularity  requests higher regularity: the three term recurrence relation is not reduced to the two term one, which in turn  request   an extra quantization of system parameter apart from the energy eigenvalue.  
   
 Notice that due to the $N, L$ dependence of $b_{0}$, the spectrum $\mathcal{E} $  is NOT linear in $N$ anymore.
Notice also that for  $a$-quantization,   $\mathcal{E} $  is linear in $N,L$ and does not depend on a quantized value of $ a_0 $ as far as $ a_0 $  is  actually given by one of those  quantized value that depends on $N$, $L$ and $b_{0}$.
  Table~\ref{tb:2} tells us all possible roots of $ a_0 $'s for each $L$ when $N=4$ and $b_{0}=1$.
  Similarly, Table~\ref{tb:3} shows us all possible roots of $ a_0 $'s for each $L$ when $N=5$ and $b_{0}=1$. As you can see easily from the table, most of the quantized values are in the linear regime where $a_{0}\approx (L+1+K)b_{0}$.
\begin{table}[!htb]
%\scriptsize
\footnotesize
\begin{center}
\begin{tabular}{|l|l|l|l|l|l|}
\hline
  & $a^{(N=4)}_{00}$ & $a^{(N=4)}_{01}$ &$a^{(N=4)}_{02}$ & $a^{(N=4)}_{03}$ & $a^{(N=4)}_{04} $ \\
\hline
\hline
L=0 & -7.50342 & -2.26852 & 2.5487  & 7.93985 & 14.2834 \\ \hline
L=1 & -9.22584 & -2.68053 & 3.72372 & 10.4374 & 17.7452 \\ \hline
L=2 & -10.4722 & -2.80774 & 4.79946 & 12.6207 & 20.8598 \\ \hline
L=3 & -11.4284 & -2.78208 & 5.84226 & 14.6311 & 23.7371 \\ \hline
L=4 & -12.1842 & -2.65493 & 6.8699  & 16.5287 & 26.4406 \\ \hline
\end{tabular}
\caption{ Roots of $ a_0 $  for $ b_0 =1$,  $N=4$. }
\label{tb:2}
\end{center}
\end{table}
\begin{table}[!htb]
\scriptsize
\begin{center}
\begin{tabular}{|l|l|l|l|l|l|l|}
\hline
  & $a^{(N=5)}_{00}$ & $a^{(N=5)}_{01}$ &$a^{(N=5)}_{02}$ & $a^{(N=5)}_{03}$ & $a^{(N=5)}_{04} $& $a^{(N=5)}_{05} $\\ \hline
\hline
L=0 & -10.5701 & -4.75187 & 0.363597 & 5.60184 & 11.6841 & 18.6724 \\ \hline
L=1 & -12.7643 & -5.82539 & 0.801156 & 7.5262  & 14.7189 & 22.5434 \\ \hline
L=2 & -14.4605 & -6.49042 & 1.30825  & 9.19107 & 17.3924 & 26.0593 \\ \hline
L=3 & -15.834  & -6.93228 & 1.86483  & 10.7358 & 19.8467 & 29.319  \\ \hline
L=4 & -16.9777 & -7.22567 & 2.45866  & 12.2089 & 22.1509 & 32.3849 \\ \hline
L=5 & -17.9475 & -7.4102  & 3.08179  & 13.6334 & 24.3443 & 35.2982 \\ \hline
\end{tabular}
\caption{  Roots of $ a_0 $  for  $ b_0 =1$, $N=5$. }
\label{tb:3}
\end{center}
\end{table}

From the explicit calculation,  we found the following pattern: List N+1 $a_{0}$ in the increasing order such that $a_{0K}$ is $K$-th one, $K=0,1,\cdots , N$.
  Then  the polynomial solution for the  $ a_{0K} $ has $K$ nodes. The number of nodes does not depend on $L$.
\begin{figure}[!htb]
%\minipage{0.24\textwidth}
 \subfigure[Polynomial $y_{4}$]
 { \includegraphics[width=0.46\linewidth]{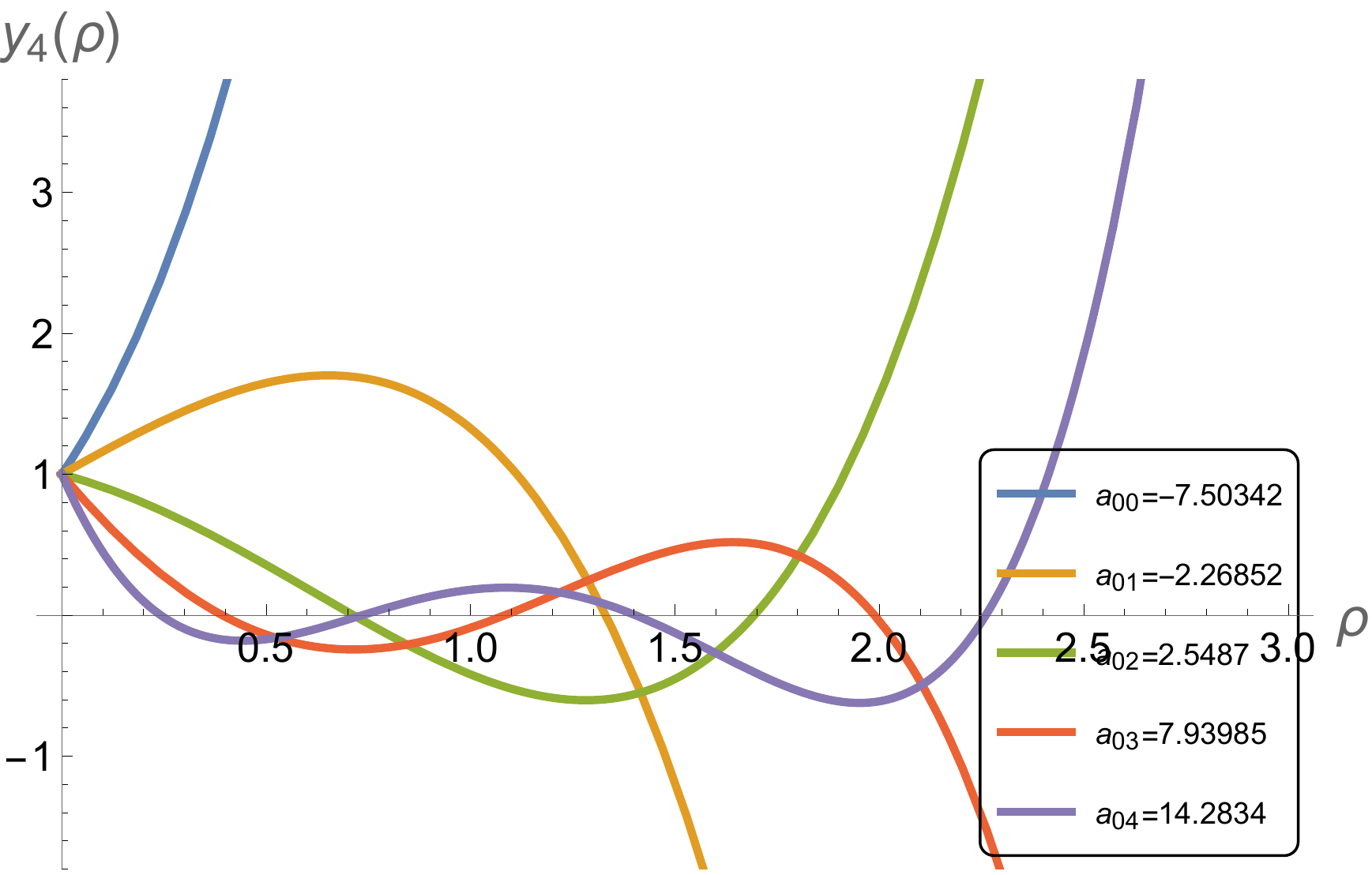}
 }
   \subfigure[Polynomial $y_{5}$]
  { \includegraphics[width=0.46\linewidth]{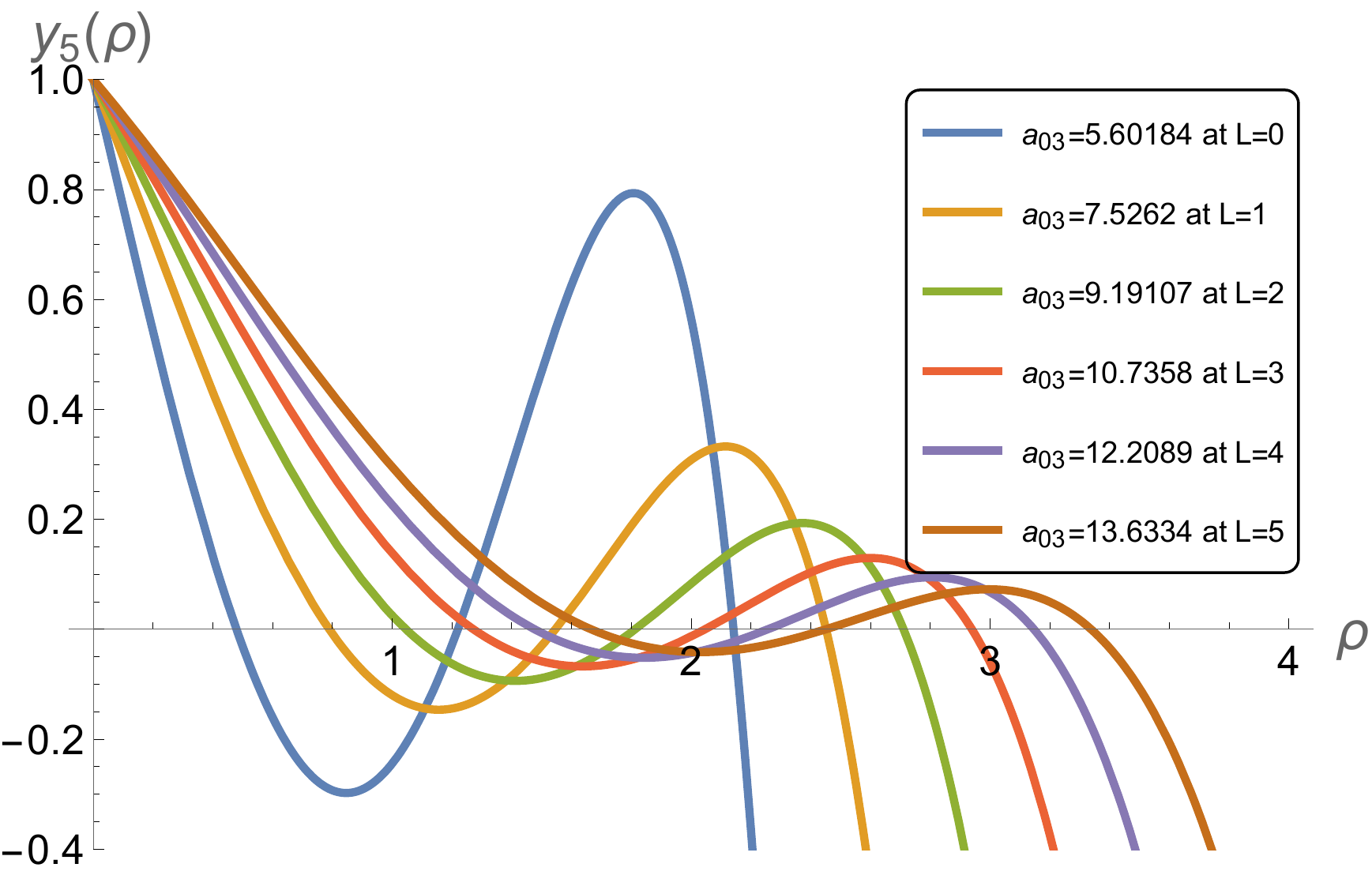}
  }
%\minipage{0.24\textwidth}
  \caption{(a)Polynomial $y_{4}$ for  various  $ a_{0K} $  
  for each $ a_{0K}$, $K=0,\cdots,4$. 
  (b)  $y_{5}$ for  various $a_{03}$ corresponding to  $L=0,1,\cdots, 5$.   }
\label{node4}
%\endminipage
\end{figure}
 Figs.~\ref{node4}(a)  shows  that polynomial $y_{4}$  with $ a_{0K}$, $K=0,1,2,3,4$ has $K$ nodes in  $N=4$ in the region $\rho>0$.  We fixed $L=0$  and $b_0 =1$.
 Figs.~\ref{node4}(b)  shows  that  polynomials $y_{5}$  with $ a_{03}$  has  3 nodes in  $N=5$   independent of the value of $L=0,1,2,3,4,5$.
There are two nodes in the unphysical region $\rho<0$.

 \vskip .3cm
%\section
{\bf 5.  Spectrum for  nonzero quark mass}: 
We  have seen that two very different models lead to the same Heun's equation.
 The spectrum of the Lichtenberg bag model for   $m_{q}=0$ was obtained in \cite{Gursey} and it is linear:
  \be
 E_{N,L}^2 = 4b_{\tau}(N+L+3/2). \label{gsey}
 \ee
 On the other hand, for $m_{q}\neq 0$,    $b$ can not be    an arbitrary value. It is  determined  by $b$-quantization because  the   parameter  $a=0$.  
 The value  of $b$ for given $N,L$ was determined numerically \cite{Bag2019} and  given by:
 
 For odd $N$,
 \be
 b_{\tau}  \approx
  \frac{8}{5}\left( N+\frac{6}{5}L+\frac{5}{3}\right)  m_{q}^2 %\left( N+ L+\frac{3}{2}\right)
 \label{qq3}
 \ee
 
 For even $N$,
 \be
 b_{\tau}  \approx
  \frac{2}{5}\left( N+\frac{6}{5}L+\frac{5}{3}\right)  m_{q}^2 %\left( N+ L+\frac{3}{2}\right)
 \label{qq4}
 \ee
 By inserting eq(\ref{qq3}) and eq(\ref{qq4}) to eq(\ref{gsey}), we see that  not only the spectrum is  highly nonlinear in $N,L$ but also the string tension is vanishingly small  in the limit of   $m_{q}\to 0$, which is   inconsistent with the nature. 
 
From the correspondence of two system given in eq(\ref{corresp}), we can read off the spectrum of holographic model from that of the bag model
by replacing 
\be
 m_{q} \to qM_0 ,  \;
  b_{\tau} /2 \to qM,   \;
  L \to |m|,  \;   {E_{N,L}^2} \to m_n^{2}.
 \ee 
Exactly parallel comment for the bag model can be applied to the spectrum. 

{\bf 6. Summary and Discussion}: 
  
In this paper, we considered  a holographic  model  where the confinement is consequence of the non-zero quark condensate $M$, so that  $M$ play the role of  the order parameter for the  confinement transition  as well as  that for  the chiral transition. 
 We also    showed    that the equation of motion of our holographic model   can be mapped to the  Lichtenberg model. 
The quark mass triggers the change of the singularity type from the hypergeometric  type to Heun's type.  The Regge trajectory  appears only at the zero  current quark mass limit   where higher singularity disappears.

Before we finish,  we discuss  a  similar model in $AdS_5$. 
The  scalar solution in AdS5 with   $m_\Phi^2=-4$ is 
\be
%\Phi =M_0 z + {M}z^3,  \hbox{ in AdS$_{5}$, } \label{Phi}
\Phi = M_0 z^{2}\ln z^{-1}+ {M}z^2,  \hbox{ in AdS$_{5}$. } \label{Phi}
\ee
If the quark mass  $M_0$ vanishes,  we still have $\Phi =  {M}z^2$, which is necessary power in $z$ to give    linear confinement. Therefore  in this model,   exactly the same calculation leads to the same  result  of  AdS4 model. 
However, there is one subtlety here.   $m_\Phi^2=2(2-4)=-4$  follows from the assumption that the dimension of  $\vev{\bar qq}$ is $\Delta=2$, while its value for the free theory is 3 for  3+1 dimensional boundary theory. 
Therefore for our senario to work, we need anomalous dimension $\gamma=-1$ 
In 2+1 dimension, on the other hand, we can simply use the $m_\Phi^2=-2$ for $\Delta=2$, which is the reason why we used $AdS_4$ model in the main text. 
 
 %\acknowledgments
\section*{Acknowledgements}
 %\begin{acknowledgements}
  We appreciate the useful discussion with  Eunseok Oh 
  and   the hospitality of the APCTP during the workshop ``Quantum Matter from the Entanglement and Holography''. 
 This  work is supported by Mid-career Researcher Program through the National Research Foundation of Korea grant No. NRF-2016R1A2B3007687.  
%\end{acknowledgements}

%\bibliographystyle{JHEP}
%\bibliography{Refs_scalar.bib}
%\bibliographystyle{model1a-num-names}

%\onecolumn

\end{document}